# Surface Tension Adjustment in a Pseudo-Potential Lattice Boltzmann Model

Anjie Hu[a], Longjian Li[a,∗], Rizwan Uddin[b]

*a. Key Laboratory of Low-grade Energy Utilization Technologies and Systems of Ministry of Education, Chongqing University, Chongqing 400030, PR China*

*b. Department of Nuclear, Plasma and Radiological Engineering, University of Illinois at Urbana Champaign, Urbana, IL 61801, USA*

**Abstract**

Pseudo-potential lattice Boltzmann models have been widely applied in many multiphase simulations. However, most of these models still suffer from some drawbacks such as spurious velocities and untunable surface tension. In this paper, we aim to discuss the surface tension of a popular pseudo-potential model proposed by Kupershtokh et al., which has attracted much attention due to its simplicity and stability. The influence of a parameter on the surface tension in the model is analyzed. Based on the analysis, we proposed a method to adjust surface tension by changing the parameter in the model. However, the density distribution and the stability of the model also depend on the parameter. To adjust the surface tension independently, the pressure tensor modifying method is introduced and numerically tested. The simulation results show that, by applying the pressure tensor modifying method, the surface tension can be adjusted with little influence on the stability and density distributions.

∗ Corresponding author. Tel.: +86 023 65111867; Fax: +86 023 65111867
*E-mail address:* longjian@cqu.edu.cn (L. Li).

# 1 Introduction

Lattice Boltzmann equation (LBE) method [1] also known as Lattice Boltzmann method (LBM) has been widely used in fluid flow simulations. Compared with the traditional fluid simulation methods, LBE method has many advantages such as mesoscopic background, strict second-order accuracy, easy for parallel computing and so on. These advantages attract many researchers to develop LBEs for different fluid flow problems. One of its promising applications is multiphase simulation. Due to its mesoscopic nature and the ability of capturing phase interface automatically, LBE method has achieved a great success in multiphase simulations. There are generally four kinds of LBE multiphase models [2] have been developed: color models [3], pseudo-potential methods [4], free energy models [5, 6] and kinetic models [7-9]. Among them, the pseudo-potential models, first proposed by Shan and Chen [4], are popularly applied because of its simplicity and stability. The phase separation of the pseudo-potential model is achieved by imposing the effect of intermolecular interactions which is represented by short-range forces between different phases.

However, the early pseudo-potential models suffer from some drawbacks, such as instability for large density ratio and untunable surface tension [10, 11]. The stability of pseudo-potential model have been highly improved in recent years, and now it can be applied in large density ratio simulation with relatively higher accuracy force schemes [10-12]. However, the surface tension adjustment of the model was rarely mentioned until recently.

Sbragaglia et al. [13] pointed out that by using high-order isotropic gradient operators, the spurious currents can be reduced and the stability of the model can be improved, hence, they developed a multi-range potential by combining the nearest-neighbor interactions and the next-nearest-neighbor interactions. The surface tension of this model can also be adjusted by changing the parameters in the model. However, it is difficult to deal with boundary condition with multi-range potential, and the parameters which adjust the surface tension also influence the density distributions [10, 13]. To adjust the surface tension independently, Li et al. [10] proposed a model by modifying the pressure tensor. Compared with Sbragaglia et al.'s model, this model is density distribution independent and has a larger surface tension adjustment range. However, since the pressure tensor modification is based on the MRT operator, it can only be applied in other operators such as LBGK by transforming from MRT operator.

In a parallel effort to improve the stability of pseudo-potential model, Kupershtokh et al. [14-16] developed a new model by combining two different force calculating models without adding next-nearest-neighbor interactions. Due to its stability and simplicity, this model has been widely used in many multiphase simulations [17-19]. However, the influence of the new force method on the surface tension has not be discussed before.

To fill the gaps, we here theoretically investigate the surface tension of the Kupershtokh et al.'s model. Based on the analysis, two surface tension adjustment approaches of this model are proposed and compared in the present work (The MRT operator is adopted here to reduce the spurious velocity and to modify the pressure tensor).

The rest paper is organized as follows. Section 2 describes the mathematical theory of Kupershtokh et al.'s model in MRT operator. Two surface tension adjustment methods of this model are discussed in section 3. Further applications of the improved model are introduced in section 4. Finally, a brief conclusion are made in section 5.

**2 Pseudo-potential model**

In the LBE method, the motion of the fluid is described by evolution of the density distribution function. The evolution equation can be written in the form of MRT operator [2] as

$$f_\alpha(\mathbf{x}+\mathbf{e}_\alpha \Delta t, t+\Delta t) - f_\alpha(\mathbf{x},t) = -\left(\mathbf{M}^{-1}\mathbf{\Lambda M}\right)_{\alpha\beta}(f_\alpha(\mathbf{x},t) - f_\alpha^{eq}(\mathbf{x},t)) + \delta_t \mathbf{F}_\alpha, \qquad (1)$$

where $f_\alpha(\mathbf{x},t)$ is the mass distribution function of particles at node $\mathbf{x}$, time $t$; $e_\alpha$ is the velocities where $\alpha = 0, 1, 2 \cdots N$; $f_\alpha^{eq}(\mathbf{x},t)$ is the equilibrium distribution. The right side of the equation is a collision operator, and $\mathbf{M}^{-1}\mathbf{\Lambda M}$ is the collision matrix, in which $\mathbf{M}$ is the orthogonal transformation matrix and $\mathbf{\Lambda}$ is a diagonal matrix which can be written in the follow form for the two-dimensional nine-velocity (D2Q9) lattice:

$$\mathbf{\Lambda} = diag(\tau_\rho^{-1}, \tau_e^{-1}, \tau_\xi^{-1}, \tau_j^{-1}, \tau_q^{-1}, \tau_j^{-1}, \tau_q^{-1}, \tau_\upsilon^{-1}, \tau_\upsilon^{-1},). \qquad (2)$$

$\mathbf{F}_\alpha$ is the force term which is given by

$$\mathbf{F}_\alpha = \mathbf{M}^{-1}\left(\mathbf{I} - \frac{1}{2}\mathbf{\Lambda}\right)\mathbf{M}\overline{\mathbf{F}}. \qquad (3)$$

Unlike lattice Bhatnagar–Gross–Krook (LBGK) operator [20], the collision of MRT operator is calculated in moment space. The density distribution function and its equilibrium distribution function can be transferred into moment space by $\mathbf{m} = \mathbf{M}\mathbf{f}$ and $\mathbf{m}^{eq} = \mathbf{M}\mathbf{f}^{eq}$. For the D2Q9 lattice, the equilibria $\mathbf{m}^{eq}$ is given by [2, 21]

$$\mathbf{m}^{eq} = \rho\left(1, -2 + 3|v|^2, 1 - 3|v|^2, v_x, -v_x, v_y, -v_y, v_x^2 - v_y^2, v_x v_y\right). \tag{4}$$

The force terms in moment space can be written as [21]

$$\tilde{\mathbf{F}} = \mathbf{M}\mathbf{F} = \left(\mathbf{I} - \frac{1}{2}\Lambda\right)\mathbf{M}\bar{\mathbf{F}} = \left(\mathbf{I} - \frac{1}{2}\Lambda\right)\mathbf{S}, \tag{5}$$

where $\mathbf{S}$ is given by

$$\mathbf{S} = \begin{bmatrix} 0 \\ 6(v_x F_x + v_y F_y) \\ -6(v_x F_x + v_y F_y) \\ F_x \\ -F_x \\ F_y \\ -F_y \\ 2(v_x F_x - v_y F_y) \\ (v_x F_x + v_y F_y) \end{bmatrix}. \tag{6}$$

In the pseudo-potential model, the interaction force can be generally calculated in two formats for nearest neighbor interaction case: The effective density type proposed by Shan and Chen [4] and the potential function type proposed by Zhang et al. [22]. The effective density model can be written as [4]

$$\mathbf{F}_1(\mathbf{x},t) = -G\psi(\mathbf{x},t)\sum_\alpha w(|\mathbf{e}_\alpha|^2)\psi(\mathbf{x}+\mathbf{e}_\alpha\delta_t,t), \tag{7}$$

where $G$ is the interaction strength, $w(|\mathbf{e}_\alpha|^2)$ are the weights, and $\psi(\mathbf{x},t)$ is effective density. The weights $w(|\mathbf{e}_\alpha|^2)$ are given by $w(1) = 1/3$ and $w(2) = 1/12$ for D2Q9 lattice. The potential function model proposed by Zhang et al. [22] can be written as

$$\mathbf{F}_2(\mathbf{x},t) = -\sum_\alpha w(|\mathbf{e}_\alpha|^2) U(\mathbf{x}+\mathbf{e}_\alpha \delta_t, t), \tag{8}$$

where $U(\mathbf{x},t)$ is potential function which is equal to $G\psi^2(\mathbf{x},t)/2$. To improve the stability of pseudo-potential model, Kupershtokh et al. [16] proposed a hybrid model by combining these two models mentioned above, which is given by

$$\mathbf{F}(\mathbf{x},t) = -A\sum_{\alpha=0}^{8} \omega_\alpha U(\mathbf{x}+\mathbf{e}_\alpha)\mathbf{e}_\alpha - (1-A)G\psi(\mathbf{x},t)\sum_{\alpha=0}^{8}\omega_\alpha \psi(\mathbf{x}+\mathbf{e}_\alpha,t)\mathbf{e}_\alpha. \tag{9}$$

In practice, both of the effective density and the potential function can be obtained by introducing a non-ideal EOS [23]:

$$p_0 = c_s^2 \rho + cG[\psi(\rho)]^2/2 = c_s^2 \rho + cU(\rho). \tag{10}$$

The stability of Kupershtokh et al.'s model is highly improved compared with the other two models [16]. However, its surface tension has not been discussed in the literatures. To analyze the surface tension of Kupershtokh et al.'s model, Taylor expansion is applied in the present work. Through Taylor expansion [24], the leading terms of the force model (Eq. (9)) can be written as

$$\mathbf{F} = -Gc^2\left[\psi\nabla\psi + \frac{e_2 c^2}{6}\psi\nabla(\nabla^2\psi) + \frac{Ae_2 c^2}{3}(\nabla^2\psi)\nabla\psi + \cdots\right]. \tag{11}$$

The corresponding pressure tensor is given by [24]

$$\mathbf{P}(\mathbf{x},t) = -A\sum_{\alpha=0}^{8}\omega_\alpha U(\mathbf{x}+\mathbf{e}_\alpha)\mathbf{e}_\alpha\mathbf{e}_\alpha - (1-A)G\psi(\mathbf{x},t)\sum_{\alpha=0}^{8}\omega_\alpha \psi(\mathbf{x}+\mathbf{e}_\alpha,t)\mathbf{e}_\alpha\mathbf{e}_\alpha. \tag{12}$$

Similarly, through Taylor expansion, the leading terms of Eq. (12) is given by

$$\mathbf{P} = \left(\rho c_s^2 + \frac{Gc^2}{2}\psi^2 + \frac{Gc^4}{12}\psi\nabla^2\psi + A\frac{Gc^4}{12}(\nabla\psi)^2\right)\mathbf{I} + \frac{Gc^4}{6}\psi\nabla\nabla\psi + A\frac{Gc^4}{6}\nabla\psi\nabla\psi + O(\partial^4) \tag{13}$$

For two dimensional problems, the elements of the pressure tensor can be written as

$$P_{xx} = \rho c_s^2 + \frac{Gc^2}{2}\psi^2 + \frac{Gc^4}{12}\left[a\left(\frac{d\psi}{dn}\right)^2 + b\psi\frac{d^2\psi}{dn^2}\right], \tag{14}$$

$$P_{yy} = \rho c_s^2 + \frac{Gc^2}{2}\psi^2 + \frac{Gc^4}{12}\left[\psi\frac{d^2\psi}{dx} + A\left(\frac{d\psi}{dx}\right)^2\right], \tag{15}$$

$$P_{xy} = P_{yx} = 0. \tag{16}$$

In the above equations, $P_{xx}$ and $P_{yy}$ are pressures normal and parallel to the interface between phases, respectively; $n$ represents the normal direction of the interface; the value of $a$ and $b$ are both related to the parameter $A$, which are given by $a = 3A, b = 3$.

Considering $d\rho/dn = 0$ in the bulk phases, the densities in the gas and liquid phases denoted by $\rho_g$ and $\rho_l$, respectively, should satisfy the follow relation [24]:

$$\int_{\rho_g}^{\rho_l}\left(p_c - \rho c_s^2 - \frac{Gc^2}{2}\psi^2\right)\frac{\psi'}{\psi^{1+\varepsilon}}d\rho = 0, \tag{17}$$

where $\varepsilon = -2a/b$. This equation is usually called the mechanical stability condition [25]. The surface tension coefficient is defined as [24]

$$\sigma \equiv \int_{-\infty}^{\infty}(P_c - P_0)dn = \int_{-\infty}^{\infty}(P_{xx} - P_{yy})dn. \tag{18}$$

According to Eq. (14-16), the above equation can be written as

$$\sigma \equiv \int_{-\infty}^{\infty}\left(\frac{Gc^4}{6}\psi\frac{d^2\psi}{dn^2} + A\frac{Gc^4}{6}\left(\frac{d\psi}{dn}\right)^2\right)dn. \tag{19}$$

Considering the boundary condition $d\psi/dx = 0$ at $x = \pm\infty$, we have

$$\sigma \equiv \int_{-\infty}^{\infty}\left((A-1)\frac{Gc^4}{6}\left(\frac{d\psi}{dn}\right)^2\right)dn. \tag{20}$$

It can be seen from Eq. (20) that the surface is linearly related to the parameter $A$ for constant density distribution. According to the previous research [14-16], the model is stable for a wide range of $A$, hence, it is possible to adjust the surface tension by changing the value of parameter $A$.

## 3 Surface tension adjust methods

According to Eq. (20), the surface tension of Kupershtokh et al.'s model can be possibly adjusted by changing the parameter $A$. Moreover, the pressure tensor of the model can also be modified for the MRT operator [10]. Based on these two ideas, two surface tension adjustment methods are proposed respectively in this section.

To numerically analyze the surface tension of present model and ensure the stability for a large range of value of the parameter $A$, here we incorporated vdW equation in the simulation, which is given by [23]

$$p = K\left(\frac{\rho RT}{1-b\rho} - a\rho^2\right), \tag{21}$$

where $a = \dfrac{27R^2T_c^2}{64p_c}$, $b = \dfrac{RT_c}{8p_c}$, [23] $K$ is applied to improve the stability of the model. When $K$ is smaller than 1, the stability of this model is highly improved, but the width of interface is also increased [26]. The parameters in the EOS are given by $a = 9/49$, $b = 2/21$, $R = 1$ [23].

### 3.1 Surface tension adjustment with parameter $A$ (method 1)

As we can see in Eq. (20), the surface tension is linearly related to $A$ for a constant density distribution, therefore a large range of surface tension could be obtained by changing parameter $A$. To verify the analysis, we first numerically invest the influence of $A$ on the surface tension.

#### 3.1.1 Influence of parameter $A$ on surface tension

To test the influence of the parameter $A$ on the surface tension, we simulated the surface tension of a stationary droplet for different values of $A$. The radius of the droplet was given by 20 lattice and was initially placed in the center of the simulation domain. The computational domain was taken as $Nx \times Ny = 100 \times 100$ lattice², and periodical boundary conditions were applied in both x- and y-directions. The parameters of the EOS (Eq. (21)) was chosen as $T = 0.9\ T_c$, $K = 0.3$ to ensure the stability of the model, and the relax time $\tau_\upsilon$ was given by 1 in the simulation. The values of $A$ were chosen from -0.9 to 0.99. The simulation surface tension was calculated by Laplace law which can be written as

$$\Delta P = P_{in} - P_{out} = \sigma / R. \tag{22}$$

The simulation results of surface tension for different $A$ values are shown in Fig. 1. It can be seen from Fig. 1 that the relationship between surface tension and the value of $A$ is almost linear, which agrees with the previous analysis. Specifically, the surface tension is equal to 0.003332 when $A = 0.9$, and it decreases to 0.000326 when $A = 0.99$. These results show that it is able to adjust the surface tension by changing the value of $A$. According to Li et al.'s research [10], the range of surface tension of multi-range pseudo-potential model [13] is in the same order of magnitude when its surface tension adjustment parameters change from 0.01 to 1. Therefore, the surface tension of present method is more adjustable compared with multi-range pseudo-potential model if we can maintain the stability of the model.

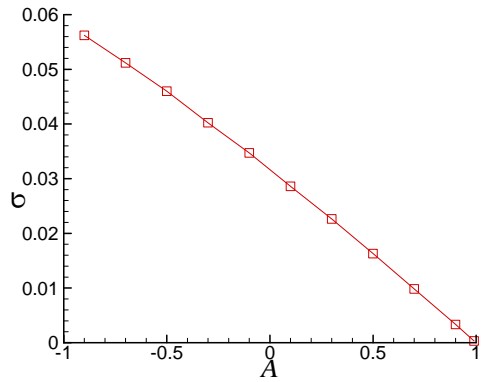

Fig.1. Surface tensions obtained for different *A* values.

**3.1.2 Drawbacks of the *A* based surface tension adjusting method**

According to the previous research [26], the parameter *A* is an important factor influencing the stability and density distribution of the model, hence, the *A* based surface tension adjustment method may influence the stability and density distribution of the model. To study these effects, we simulated the density distribution of plane two-phase interface with different values of the parameter *A*.

The parameters were the same as the previous section except we initially give a straight interface instead of a circle droplet. The simulation results are shown in Fig. 2. It can be seen from Fig. 2 that both liquid density and gas density change with the parameter *A*, and the density of gas changed about 25% when *A* changed from 1 to -0.9. These results show that the surface tension cannot be adjusted independently of the density distribution.

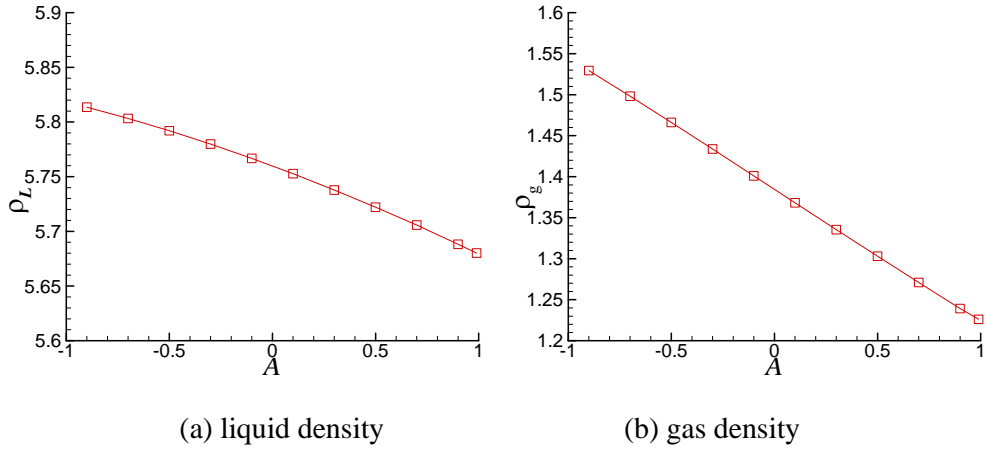

Fig. 2. Coexistence densities for different *A* values.

Stability of the model with different *A* values was also numerically investigated in the present work. Generally, the lower the temperature is, the larger the density ratio is and the more difficult for the model to stay stable, hence, we simulated the lowest temperature this model can get when $A = 0.5$ and $A = -0.5$, respectively. Simulation results show that: when $A = -0.5$, the lowest temperature can be simulated is about $0.69T_c$, and the corresponding density ratio is 22.9; when $A = 0.5$, the lowest temperature is about $0.79T_c$, and the corresponding density ratio is 16. Hence, a conclusion can be drawn that the value of the parameter *A* has significance influence on the stability of the model.

According to the above analysis, the parameter *A* based surface tension adjustment method may lead to stability problem and cause obvious density change in the simulation results. Since the stability of the model is sensitive for large density ratios, this method can only be applied when the density ratio is relatively small.

### 3.2 Pressure tensor modifying method (method 2)

Since the surface tension of the pseudo-potential model is determined by the difference between the pressures normal and parallel to the interface, it is possible to adjust the surface tension by modifying the pressure tensor. In this section, the approach of modifying pressure tensor for Kupershtokh et al.'s model will be proposed and numerically discussed.

### 3.2.1 Basic theory

It is easy and effective to adjust surface tension by changing the parameter $A$, especially for LBGK model. However, this approach is limited in low density ratios and it influences the density distribution of the model. The reason for this problem is that both pressures normal (Eq. (14)) and parallel (Eq. (15)) to the interface are changed. Since the surface tension is determined by the difference between pressures normal and parallel to the interface and the density distribution is determined by the mechanical stability condition which is only related to the normal pressure (Eq. (14)), to adjust the surface tension independently, we need to modify the parallel pressure without changing the normal pressure. To achieve this goal, Li et al.'s [10] pressure tensor modifying method is introduced in the present work.

Here we introduce a new tensor

$$\mathbf{Q} = \kappa \left[ (1-A)\psi(\mathbf{x}) \sum_{\alpha=1}^{N} w(|\mathbf{e}_\alpha|^2)(\psi(\mathbf{x}+\mathbf{e}_\alpha)-\psi(\mathbf{x}))\mathbf{e}_\alpha \mathbf{e}_\alpha + \frac{1}{2}A\sum_{\alpha=1}^{N} w(|\mathbf{e}_\alpha|^2)(\psi(\mathbf{x}+\mathbf{e}_\alpha)^2-\psi(\mathbf{x})^2)\mathbf{e}_\alpha \mathbf{e}_\alpha \right]$$

. (23)

The leading order of above tensor is given by

$$\mathbf{Q} = \kappa\left(\left(\frac{Gc^4}{12}\psi\nabla^2\psi + A\frac{Gc^4}{12}(\nabla\psi)^2\right)\mathbf{I} + \frac{Gc^4}{6}\psi\nabla\nabla\psi + A\frac{Gc^4}{6}(\nabla\psi)^2\right). \tag{24}$$

Note that when $A$ is equal to 0, this tensor is the same with Li et al.'s method, and $(Q_{xx} - Q_{yy})$ is equal to $(P_{xx} - P_{yy})$ which directly related to the surface tension, we add this term to the Navier-Stokes equation by modifying the MRT LBE. The modified MRT LBE is given by

$$\mathbf{m}^* = \mathbf{m} - \Lambda(\mathbf{m} - \mathbf{m}^{eq}) + \delta_t\left(\mathbf{I} - \frac{\Lambda}{2}\right)\mathbf{S} + \delta_t\mathbf{C}, \tag{25}$$

where the source term $\mathbf{C}$ is given by

$$\mathbf{C} = \begin{bmatrix} 0 \\ 1.5\tau_e^{-1}(Q_{xx} + Q_{yy}) \\ 0 \\ 0 \\ 0 \\ 0 \\ 0 \\ -1.5\tau^{-1}(Q_{xx} - Q_{yy}) \\ -\tau_v^{-1}Q_{xy} \end{bmatrix}. \tag{26}$$

Note that the third element of $\mathbf{C}$ given by Li et al. [10] is $-1.5\tau_e^{-1}(Q_{xx} + Q_{yy})$ instead of 0, since this term has no influence on the N-S equations, it is given by 0 in the present work. Through the Chapman-Enskog analysis, the new N-S equation is given by [27]

$$N-S_{new} = N-S_{orginal} - \kappa\left(\left(\frac{Gc^4}{6}\psi\nabla^2\psi + A\frac{Gc^4}{6}(\nabla\psi)^2\right)\mathbf{I} - \frac{Gc^4}{6}\psi\nabla\nabla\psi - A\frac{Gc^4}{6}(\nabla\psi)^2\right),$$

$$\tag{27}$$

and the pressure tensor of the new N-S equation is given by

$$\mathbf{P} = \left( \rho c_s^2 + \frac{Gc^2}{2}\psi^2 + \frac{Gc^4}{12}(1+2k)\psi\nabla^2\psi + A(1+2k)\frac{Gc^4}{12}(\nabla\psi)^2 \right)\mathbf{I}$$

$$+ (1-\kappa)\left( \frac{Gc^4}{6}\psi\nabla\nabla\psi + A\frac{Gc^4}{6}\nabla\psi\nabla\psi \right) + O(\partial^4). \tag{28}$$

It can be seen from the above equation that $P_{xx}$ is independent of $\kappa$, and the surface tension is determined by $(1-\kappa)\left( \frac{Gc^4}{6}\psi\nabla\nabla\psi + A\frac{Gc^4}{6}\nabla\psi\nabla\psi \right)$. When the value of $\kappa$ changes from 0 to 1, the surface tension will be linearly reduced to zero.

### 3.2.2 Numerical test of method 2

For the sake of comparing these two surface tension adjustment methods, we tested the pressure tensor modifying method with the same EOS and parameters. In these cases, *A* is equal to -0.5 to ensure the stability of the model. The value of parameter $\kappa$ is changed from 0 to 0.99, and the simulation results are presented in Fig. 3. It can be seen from Fig. 3 that the relationship between parameter $\kappa$ and surface tension is totally linear. Specifically, the surface tensions are 0.0459, 0.00448, 0.000452, respectively when the value of $\kappa$ are equal to 0, 0.9, 0.99. These results agree with the theory analysis, hence, a conclusion can be drown that the proposed method is capable to obtain a large range of surface tension of the model.

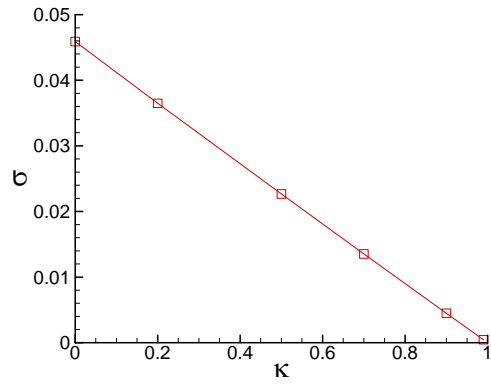

Fig. 3. Surface tensions for different $\kappa$ values.

To further analysis the influence of the additional pressure tensor, we simulated the density distribution for different values of $\kappa$ for the straight interface case. The simulation results are shown in Table 1. It can be seen from this table that both liquid densities and gas densities for different $\kappa$ are absolutely constant. These results suggest that the modified pressure tensor has no influence on the mechanical solution of the model, hence, it can adjust the surface tension independently of the density distribution.

Table 1. The coexistence densities for different $\kappa$ values.

| $\kappa$ | Rho_l | Rho_g |
|---|---|---|
| 0 | 5.79198 | 1.46621 |
| 0.2 | 5.79198 | 1.46621 |
| 0.5 | 5.79198 | 1.46621 |
| 0.7 | 5.79198 | 1.46621 |
| 0.9 | 5.79198 | 1.46621 |
| 0.99 | 5.79198 | 1.46621 |

We further studied the influence of $\kappa$ on the stability of this model. Table 2 shows the lowest temperatures can be simulated with the model for different values of $\kappa$ and $A$. It can be seen from this table that the lowest temperatures are the same for different $\kappa$ values, so a conclusion can be drawn that the stability of the model is not influenced.

Table 2. The lowest temperature can be achieved for different $\kappa$ and $A$ values.

| $\kappa$ | T ($A$=0.5) | T ($A$=−0.5) |
| --- | --- | --- |
| 0 | 0.69Tc | 0.8Tc |
| 0.2 | 0.69Tc | 0.8Tc |
| 0.5 | 0.69Tc | 0.8Tc |
| 0.7 | 0.69Tc | 0.8Tc |
| 0.9 | 0.69Tc | 0.8Tc |
| 0.99 | 0.69Tc | 0.8Tc |

**3.3 summaries**

In this section, two surface tension adjustment methods were proposed for Kupershtokh et al.'s pseudo-potential LBE model based on the theory analysis of its pressure tensor. Comparing with method 2, method 1 is easier to be applied in LBGK operator. However, method 1 cannot give a constant density ratio and ensure the stability, so we proposed method 2 to solve these problems in MRT operator. Due to it better stability and accuracy, in the next section, method 2 will be further discussed in the case of larger density ratio and some standard simulation problems.

**4 Application of method 2**

As we can see in the previous section, the density ratio of the model cannot get too large when vdW Equation is applied. To apply this method in larger density ratio, the p-r EOS is applied in this section, which is given by [23]

$$p = K\left(\frac{\rho RT}{1-b\rho} - \frac{a\alpha(T)\rho^2}{1+2b\rho-b^2\rho^2}\right), \tag{29}$$

where $\alpha(T) = [1+(0.37464+1.54226\omega_0-0.26992\omega_0^2)\times(1-\sqrt{T/T_c})]^2$, $a = 0.45724 R^2 T_c^2 / p_c$, $b = 0.0778 RT_c / p_c$ [23]. The parameters are chosen as $b = 2/21$, $R = 1$, $\omega_0 = 0.344$. In present research, $T$ is chosen as $0.8T_c$, to ensure the stability of the model, $A = -0.3$, and $K = 0.5$. The corresponding density ratio is 430, and the interface width is about 4 lattices.

**4.1 Laplace law**

First we investigated the relationship between pressure difference and the radius of the droplet. According to Eq. (22), if the surface tension is constant, the pressure difference $\delta P$ between inside and outside bubble should be proportional to the reciprocal of the radius $R$. To verify this model, we simulated the pressure difference for different radius and different surface tensions. The parameter $\kappa$ in Eq. (23) was chosen as 0, 0.9, and 0.99, respectively. Initially, a single droplet was placed in the center of the simulation domain of $120\times120$ lattice². The initial droplet radius varies from 20 to 40 in the simulations. Figure 4 shows the simulation results of the relationship between the pressures and radius. To clearly compare the results under different surface tensions, logarithmic coordinate is applied for pressure difference. It can be seen from Fig. 4 that $\delta P$ is proportional to the reciprocal of radius. However, slight differences appear when $\kappa = 0.99$. The reason of this error may be that the error terms in N-S equation cannot be neglected when the surface tension is too small.

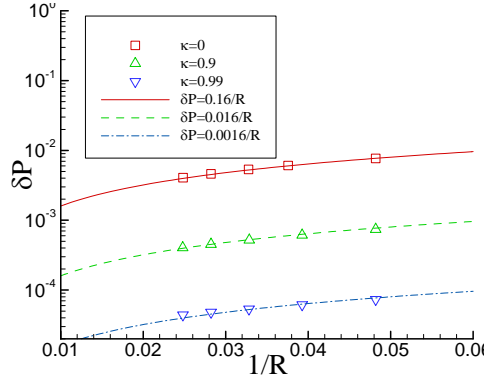

Fig. 4. Numerical validation of the Laplace's law for different values of

### 4.2 Droplet Oscillation

The liquid cylinder oscillations phenomenon is also considered in present work. Oscillatory behavior exhibits when the liquid cylinder is slightly perturbed from its equilibrium circular shape. The oscillation period of this phenomenon is given by [28]

$$T_a = \left[ 2\pi n(n^2 - 1)\frac{\sigma}{\rho_l R_0^3} \right]^{-0.5}, \tag{30}$$

where $T_a$ is the oscillation period, $\sigma$ is the surface tension, $\rho_l$ is the liquid density, and $R_0$ is the equilibrium radius of the drop. $n$ denotes the oscillatory mode, which is 2 for initial elliptical shape [21]. It should be noted that the effect of the surrounding gas density and viscosity are not considered in Eq. (30).

The simulation domain was chosen as $120 \times 120$ lattice$^2$ domain, and initially, an elliptic droplet was located at the center of the computational domain. The major radius and the minor radius of the drop were given by $R_{max} = 30$, $R_{min} = 27$, respectively. The equilibrium droplet radius was given by $R_0 = \sqrt{R_{max} R_{min}}$. The values of surface tension parameter $\kappa$ were chosen as -1, 0, 0.5 in the simulation, the corresponding surface tensions were 0.32, 0.16, 0.08, and the oscillation periods given by [Eq](). (30) were 1698, 2402, 3397, respectively. The viscosities of gas phase were 0.2 for all cases, and liquid viscosities $\upsilon_l$ are 0.1 and 0.05 for each $\kappa$.

[Figure 5]() shows the oscillatory when $\kappa = -1$. The corresponding surface tension is 0.32. It can be seen that the liquid viscosities have little influence on the oscillatory period. When $\upsilon_l = 0.05$, the period is 1639, the corresponding error is 3.5%; when $\upsilon_l = 0.1$, the period is 1602, the corresponding error is 5.6%.

[Figure 6]() shows the oscillatory when $\kappa = 0$. The influence of liquid viscosities on the oscillatory is also not obvious. When $\upsilon_l = 0.05$, the period is 2266, the corresponding error is 5.7%; when $\upsilon_l = 0.1$, the period is 1602, the corresponding error is 7.4%.

[Figure 7]() shows the oscillatory when $\kappa = 0.5$. The influence of liquid viscosities on the oscillatory is obvious compared with the previous two cases. When $\upsilon_l = 0.05$, the period is 2266, the corresponding error is 6.6%; when $\upsilon_l = 0.1$, the period is 3174, the corresponding error is 15.9%.

According to the above results, the viscosity of fluid has little influence on the oscillatory period when the surface tension is relatively large. However, when the surface tension is small, the influence of viscosity becomes large, and the increased viscosity causes the dumping of the oscillatory and reduces the period in the model.

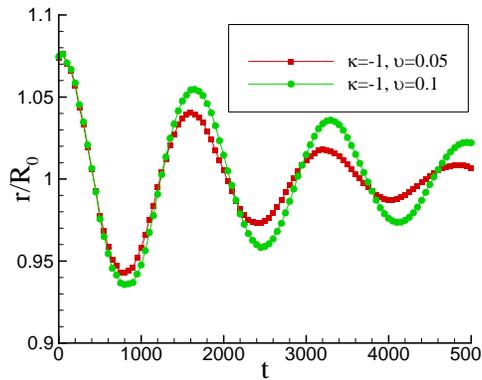

Fig. 5. Oscillation of an infinitely long liquid cylinder initially with an elliptical cross-section, at two different liquid viscosities; $\sigma = 0.32$.

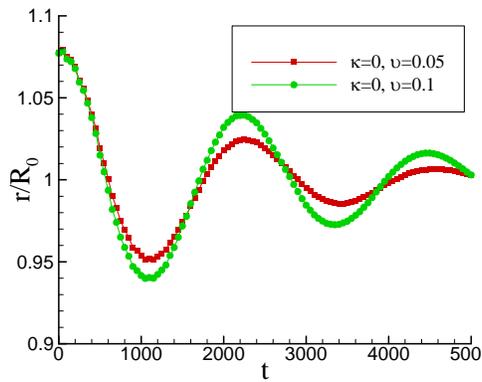

Fig. 6. Oscillation of an infinitely long liquid cylinder initially with an elliptical cross-section, at two different liquid viscosities; $\sigma = 0.16$.

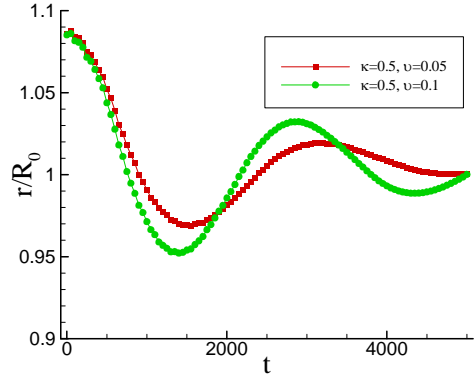

Fig. 7. Oscillation of an infinitely long liquid cylinder initially with an elliptical cross-section, at two different liquid viscosities; $\sigma = 0.08$.

## 5. Conclusions

In this paper, theoretical and numerical analysis were adopted for Kupershtokh et al.'s model. Two surface tension adjustment methods were proposed based on the analysis. The first one is adjusting its surface tension by changing the exited parameter *A*, simulation results show that it can get a large range of surface tension and be easily applied in both LBGK and MRT operators. However, this parameter is density depended and has influence on the stability of the model, hence, it can only be applied when the density ratio is relatively small. The second method is based on the idea of modifying pressure tensor. By introducing a modifying tensor, the surface tension can be adjusted independently of the density distributions and the stability of the model. It has also been validated through numerical simulations of Laplace law and capillary waves. These works provide some ways to adjust the surface tension in Kupershtokh et al.'s pseudo-potential LBE model, it may be useful for further application of the pseudo-potential LB model.

**Acknowledgements**

This work is sponsored by National Natural Science Foundation of China (51076172) and Science and Technology Innovation Foundation of Chongqing University (CDJXS11142232).